%% file: draft_JHEP.tex
\documentclass[a4paper,11pt]{article}

\usepackage{jheppub}
\usepackage[T1]{fontenc}
\usepackage[utf8]{inputenc}
\usepackage[american]{babel}
\usepackage{csquotes}
\usepackage{subcaption}
\usepackage{float}

\usepackage{lipsum}
\usepackage{graphicx}
\usepackage{dcolumn}
\usepackage{tabularx}
\usepackage{bm}
\usepackage{amssymb}
\usepackage{amsmath}
\usepackage{booktabs}
\usepackage{multirow}
\usepackage{makecell}
\usepackage[colorlinks=true]{hyperref}
\usepackage{lineno}
\usepackage{soul}

\title{\boldmath {Search for Double Beta Decay of $^{136}$Xe to the $0^+_1$ Excited State of $^{136}$Ba with PandaX-4T}}

\collaborationImg{\includegraphics[width=0.3\textwidth]{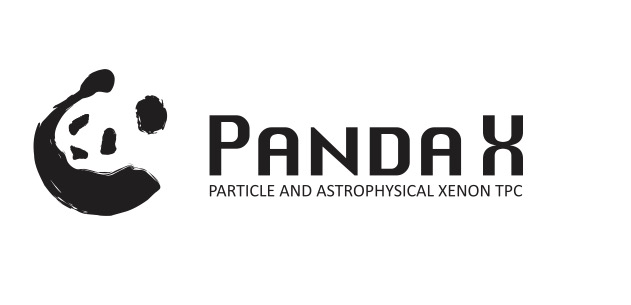}}
\input{authorlist1002.tex}

\date{\today}
\abstract{
We perform a search of double beta decay of $^{136}$Xe to the excited state, $0^+_1$, of $^{136}$Ba (2$\nu \beta \beta$-0$_1^+$), using the dual-phase xenon detector of PandaX-4T with the first 94.9-day commissioning data.
The multi-site events are reconstructed up to the MeV energy scale, which helps to improve the background model significantly. 
The background contribution from the stainless steel platform outside PandaX-4T cryostat is evaluated for the first time.
No significant evidence for 2$\nu \beta \beta$-$0_1^+$ is observed, resulting in a lower limit on half-life of $7.5 \times 10^{22}$~yr at the 90\% confidence level. This is the first experimental limit on such a rare decay in a natural xenon-based detector.
}
\begin{document}

\maketitle
\flushbottom

\section{Introduction}
\label{sec:intro}
Double beta decay is an established second-order process in the standard model of the electroweak interaction~\cite{Goeppert-Mayer:1935uil, Saakyan:2013yna}, in which two electrons and two antineutrinos are emitted simultaneously in a nucleus. If neutrinos are Majorana particles~\cite{Furry:1939qr}, neutrinoless double beta decay~(0$\nu \beta \beta$)~\cite{Avignone:2007fu} could happen without antineutrinos emitted. The observation of 0$\nu \beta \beta$ would demonstrate the Majorana nature of neutrinos.

Double beta decay to the ground state of daughter nucleus~(2$\nu \beta \beta$) has been directly observed in eleven isotopes with half-lives exceeding 10$^{18}$~yr, including $^{136}$Xe, $^{130}$Te, and $^{76}$Ge,  etc~\cite{KamLAND-Zen:2019imh, NEXT:2021dqj, PandaX:2022kwg,GERDA:2023wbr, Majorana:2022udl}. 
Most of these isotopes are used to search for 0$\nu \beta \beta$ in experiments. The current limits on 0$\nu \beta \beta$ half-life time are at 10$^{26}$~yr level~\cite{CUORE:2022jto, KamLAND-Zen:2022tow, Garfagnini:2024rvs}. 
Double beta decay transitions to excited states of the daughter nucleus, on the other hand, are allowed process in the standard model, but are significantly rarer in comparison to 2$\nu \beta \beta$  to the ground state due to smaller transition energies~\cite{Jokiniemi:2022yfr, Barabash:2017bgb}. 
However, measurements of different transitions provide valuable information for the development of theoretical schemes to calculate the relevant nuclear matrix elements~\cite{Barea:2015kwa}. 
Understanding the nuclear matrix elements is crucial for studying the nuclear structure and nuclear interactions related to 0$\nu \beta \beta$.

Double beta decay to the first $0^+$ excited state of the daughter nucleus, denoted hereafter as $0_1^+$ (2$\nu \beta \beta$-0$_1^+$), is considered as the next most accessible decay~\cite{Barabash:2017bgb, KamLAND-Zen:2015tnh}. 
\begin{figure}[htbp]
  \centering
  \includegraphics[width=\columnwidth]{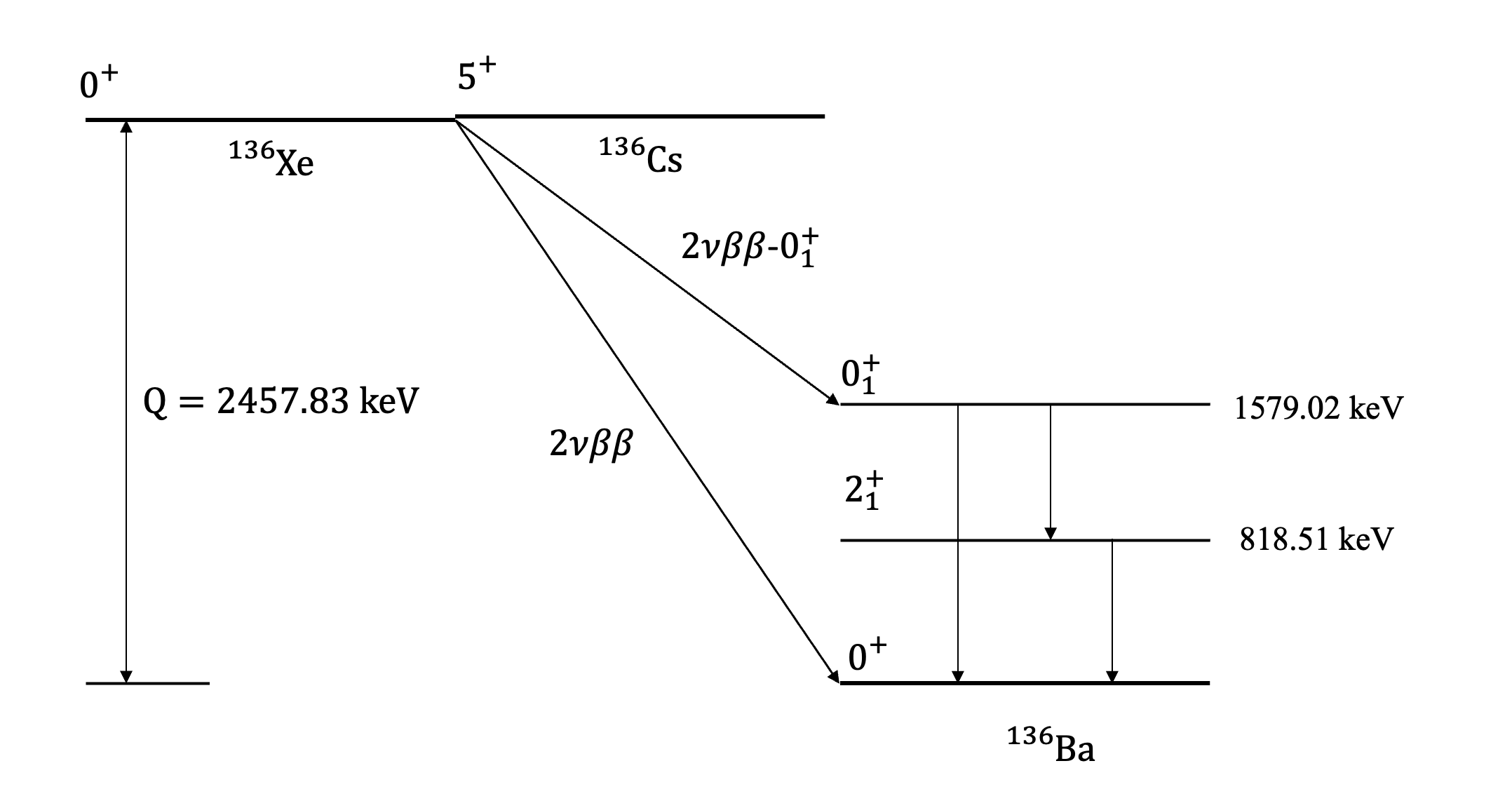}
  \caption{Decay scheme of $^{136}$Xe double beta decay ~(2$\nu \beta \beta$) and double beta decay to the excited states~(2$\nu \beta \beta$-0$_1^+$).}
  \label{fig:decay}
\end{figure}
These decays are typically accompanied by the emission of de-excitation $\gamma$ rays, 
producing a distinct signature that differentiates them from typical $\gamma$ backgrounds and 2$\nu \beta \beta$ events. 2$\nu \beta \beta$-0$_1^+$ was first observed in $^{100}$Mo~\cite{Barabash:1995fn} in 1995, 
and then its half-life was precisely measured by the NEMO-3 experiment~\cite{NEMO-3:2014pkc} where the tracks of all $\beta$ and $\gamma$ particles were detected. 
In 2014, 2$\nu \beta \beta$-0$_1^+$ of $^{150}$Nd was observed through the measurement of $\gamma$ coincidence~\cite{Kidd:2014hra}. 
CUORE, MAJORANA, and KamLAND-Zen experiments have set half-life limits on 2$\nu \beta \beta$-0$_1^+$ at the 10$^{23}$yr level for $^{130}$Te, $^{76}$Ge, and $^{136}$Xe, respectively~\cite{CUORE:2021xns, MAJORANA:2020shy, KamLAND-Zen:2015tnh}. 
The $^{136}$Xe 2$\nu \beta \beta$-0$_1^+$ decay, with a Q value of 878.8~keV, corresponds to $^{136}$Xe decaying into the the $0^+_1$ excited state of $^{136}$Ba.
As shown in Figure~\ref{fig:decay}, the energies of the de-excitation $\gamma$ rays are either a cascade of 760.5 keV and 818.5 keV, or a single 1579.0 keV. 
The expected half-life in theory for the  $^{136}$Xe 2$\nu \beta \beta$-0$_1^+$ ranges from 10$^{22}$ to 10$^{26}$ yr, depending on the model used~\cite{Jokiniemi:2022yfr}. 
Current experiments are entering the region of the predictions, thus 2$\nu \beta \beta$-0$_1^+$ of $^{136}$Xe is considered as one of the most promising rare processes to be discovered in the future. 
More recently, the EXO-200 experiment has performed a search using a total isotope exposure of 234.1 kg$\cdot$yr, establishing the best experimental limit on half-life at 1.4$\times$10$^{24}$~yr~\cite{EXO-200:2023pdl}.

The PandaX-4T experiment uses a dual-phase time projection chamber with 3.7-tonne natural xenon in the sensitive volume to search for Weakly Interacting Massive Particles dark matter~\cite{PandaX-4T:2021bab} and other rare events~\cite{PandaX:2023ggs, PandaX:2024pme, PandaX:2024jjs, PandaX:2023xgl}. 
The detector can record both the three-dimensional position and energy deposition of an event inside. 
The PandaX-4T detector contains approximately 300~kg of $^{136}$Xe within the sensitive volume.
Recently, PandaX-4T has developed a dedicated analysis pipeline for the MeV energy scale signals and performed calibration campaigns up to the MeV energy range, allowing the study of double beta decay. 
In previous analysis~\cite{PandaX:2022kwg, PandaX:2023ggs}, we have searched for $^{136}$Xe 2$\nu \beta \beta$ events which are primarily single-site~(SS) events. In contrast, the MeV-scale $\gamma$ background events can scatter multiple times,  leading to multi-site~(MS) events. 
In this work, we focus on the search for $^{136}$Xe 2$\nu \beta \beta$-0$_1^+$. Most signals associated with $\gamma$s are MS events, therefore we extended the signal selections to MS events.
The combination of SS and MS events allows us not only to make a significant improvement to our background model but also to derive a combined limit for $^{136}$Xe 2$\nu \beta \beta$-0$_1^+$.

This paper is structured as follows: Section~\ref{sec:p4} describes the PandaX-4T detector and its working principle; Section~\ref{sec:data} discusses the data production, event selection procedures, and event reconstruction; Section~\ref{sec:bkg} presents the background components and their Monte Carlo (MC) simulations; Section~\ref{sec:fit method} explains the contributions of systematic uncertainties and the fitting method used in this analysis; Section~\ref{sec:bkg fit} demonstrates the background-only fits and a highly precise fitted background model; Section~\ref{sec:results} reports the half-life limit of the $^{136}$Xe 2$\nu \beta \beta$-0$_1^+$; and a summary is provided in Section~\ref{sec:conclusion}.

\section{The PandaX-4T experiment}
\label{sec:p4}

\begin{figure}[htbp]
  \centering
  \includegraphics[width=\columnwidth]{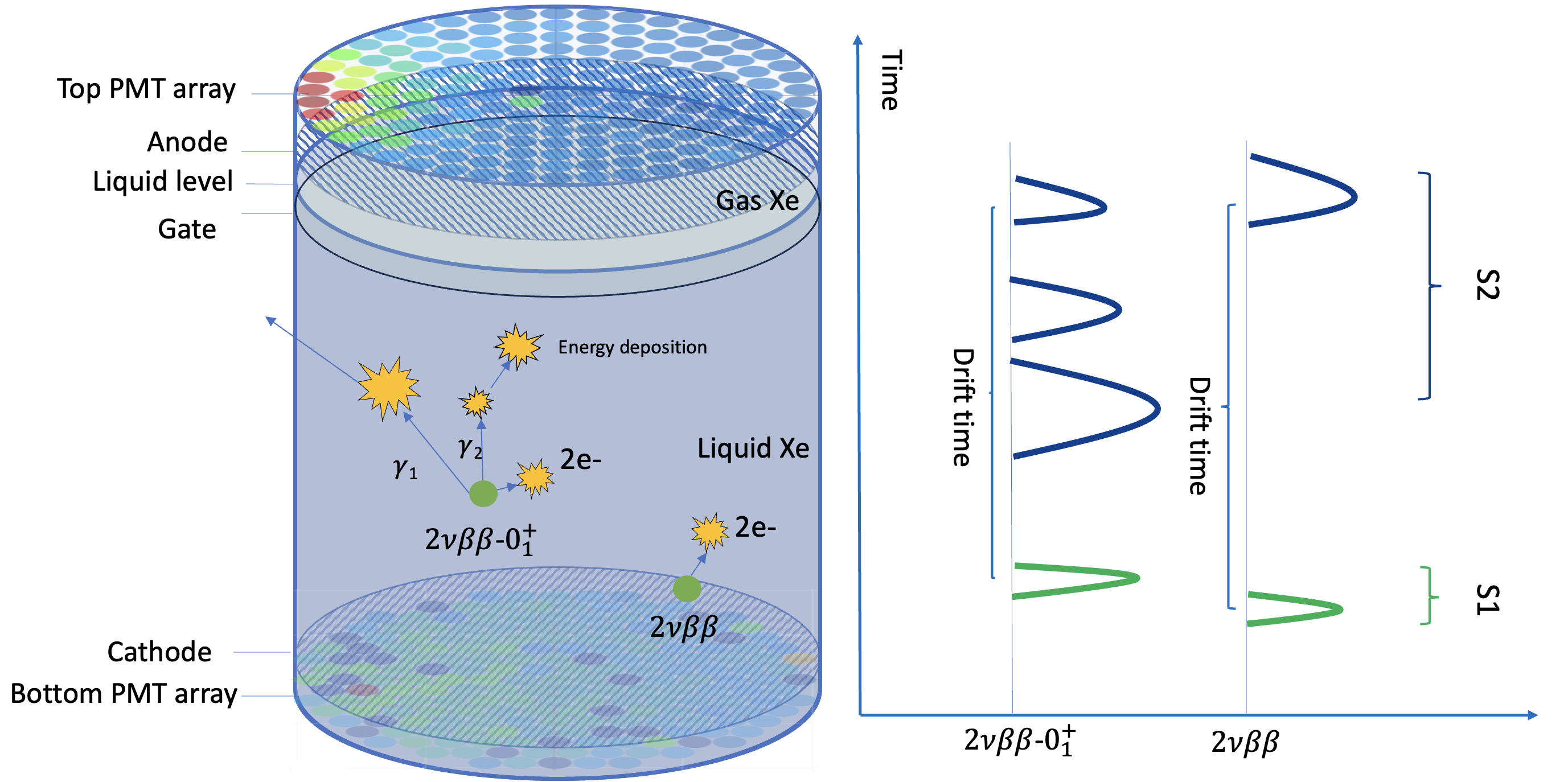}
  \caption{Schematic diagram of the time projection chamber~(left) and typical waveforms for $^{136}$Xe 2$\nu \beta \beta$-0$_1^+$ and 2$\nu \beta \beta$~(right).}
  \label{fig:TPC}
\end{figure}

PandaX-4T is the third generation PandaX experiment, located in the B2 hall of China Jinping Underground Laboratory~\cite{Li:2014rca}. 
The detector is a cylindrical dual-phase time projection chamber with a diameter and height of 118.5 cm, filled with 3.7 tonnes of natural xenon in the sensitive volume. The sensitive volume is surrounded by the field cage, with the gate electrode at the top and the cathode plane at the bottom.
 As illustrated in Figure~\ref{fig:TPC} (left), 
Photomultiplier tubes~(PMT) arrays, consisting of 199 and 169 three-inch Hamamatsu PMTs, are installed below the cathode and above the anode for signal readout. 
A detailed description of the detector is provided in Ref.~\cite{PandaX-4T:2021lbm}.

The detector records both the three-dimensional position and the energy deposition of an event inside the sensitive volume. 
An energetic event deposits energy inside the liquid xenon through scintillation light (S$1$) and ionized electrons. 
These electrons drift towards the gas-liquid interface under the effect of the electric field, and then a delayed electroluminescence signal~(S$2$) is generated by a strong electric field in the gas phase. 
The S$1$ and S$2$ signals are collected by the top and bottom PMT arrays. 
We will use the subscripts "T" and "B" to represent the S$1$ and S$2$ received from the top and bottom PMT arrays, respectively.
The total energy of an event is determined by combining the area of S$1$ and S${2_\text{B}}$ waveform.
The event position in the horizontal X-Y plane is reconstructed using the distribution of S$2$ detected by each PMT in the top array.
The vertical position Z of an event is determined based on the drift velocity of electrons and the time gap between S$1$ and S$2$. 
The detector is capable of identifying MS and SS events by counting the number of observed S$2$ peaks~\cite{PandaX:2022kwg}.
As shown in Figure~\ref{fig:TPC}, the 2$\nu \beta \beta$-0$_1^+$ decay typically deposits energy as an MS event, characterized by multiple S$2$ peaks, whereas most 2$\nu \beta \beta$ events manifest as SS events.

\section{Data production and event reconstruction}
\label{sec:data}

A total of 94.9 days of data were collected during the commissioning run, which spanned from November 28, 2020, to April 16, 2021, as detailed in Ref.~\cite{PandaX-4T:2021bab}. 
The data production and event selection procedures follow those outlined in Refs.~\cite{PandaX:2023ggs, PandaX:2024kjp, PandaX:2024xpq}, where the SS event reconstruction in the energy range between 25~keV and 2.8~MeV has been successfully achieved.
In this study, we focus on the reconstruction of MS events, including SS/MS discrimination, event selection, event reconstruction, and detector response.
Both constructed energy spectra for MS and SS are selected to search for 2$\nu \beta \beta$-0$_1^+$.

\subsection{MS and SS discrimination}

The discrimination between SS and MS events is performed by counting the number of peaks in the S${2_\text{B}}$ waveform~\cite{PandaX:2022kwg}. Most of the 2$\nu \beta \beta$-0$_1^+$ events are MS events, while the dominant background consists of 2$\nu \beta \beta$ SS events. The temporal profile of the summed S${2_\text{B}}$ waveform provides a signature to distinguish between MS and SS events. Due to the longitudinal diffusion of the ionized electrons, the S${2_\text{B}}$ signal is broadened. A dynamic time window is applied to count S${2_\text{B}}$ peaks, which is based on the drift distance. The width of this dynamic window is determined by the 1$\sigma$ width of the highest-smoothed S${2_\text{B}}$ waveform, ranging from $\pm$0.32 to $\pm$3.2~$\mu$s. After identifying the S${2_\text{B}}$ peaks, events with a single peak are classified as SS events, while those with multiple peaks are classified as MS events~\cite{PandaX:2023ggs}.

 \begin{figure}[htbp]
  \centering
  \includegraphics[width=\columnwidth]{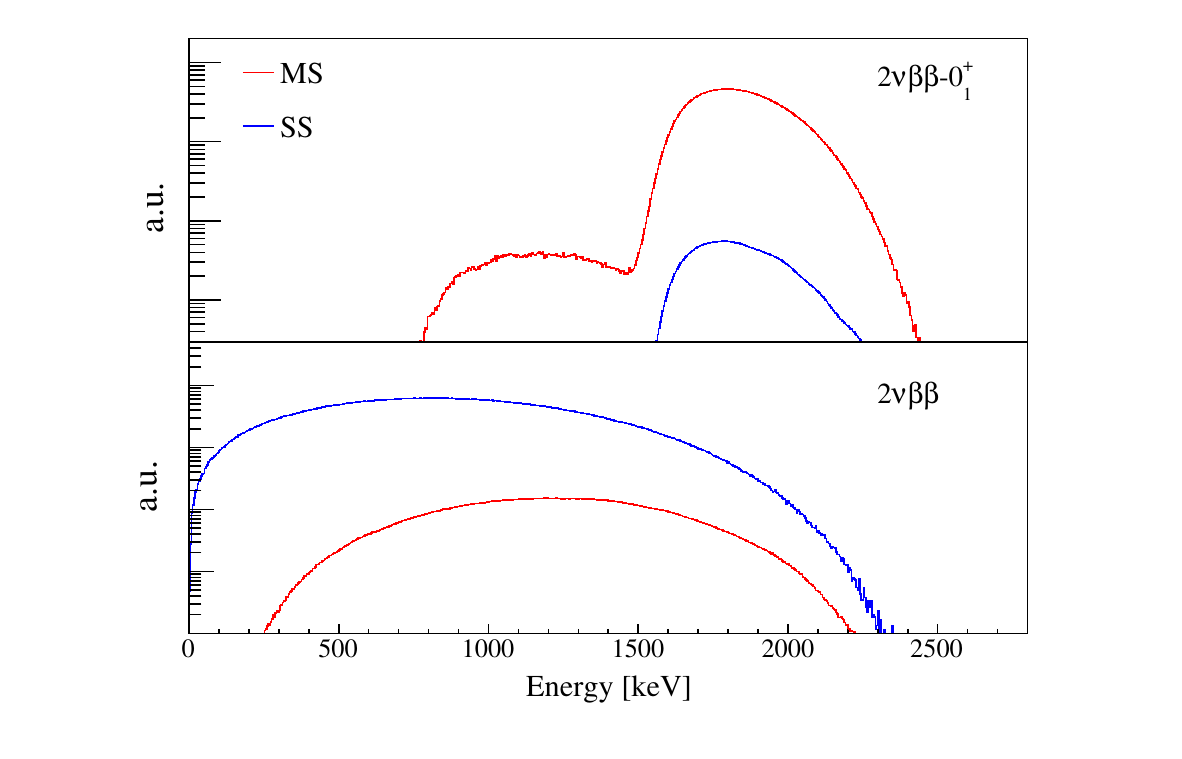}
  \caption{Energy spectra of $2\nu \beta \beta$-$0_1^+$ (top) and $2\nu \beta \beta$ (bottom) in FV with detector response. The discrimination between MS and SS events is effective for the selection of $2\nu \beta \beta$-$0_1^+$ events. }
  \label{fig:energy spe}
\end{figure}

The MS (SS) ratio is defined as the number of MS (SS) events divided by the total number of events, expressed as MS/(SS+MS) for MS events and SS/(SS+MS) for SS events.
Based on MC simulations, the MS ratio for $2\nu \beta \beta$-$0_1^+$ is 98.8\% within the region  of interest~(ROI), while the MS ratio for $2\nu \beta \beta$ is 4.3\%. 
These simulations are conducted using BambooMC, a Geant4-based~\cite{GEANT4:2002zbu} MC framework developed by the PandaX collaboration~\cite{Chen:2021asx}. 
The $^{136}$Xe $2\nu \beta \beta$-$0_1^+$ and $2\nu \beta \beta$ events have been simulated using the Decay0 generator toolkit~\cite{Ponkratenko:2000um}, which is based on theoretical calculations.
Figure~\ref{fig:energy spe} presents the simulated total deposited energy spectra for $^{136}$Xe $2\nu \beta \beta$-$0_1^+$ and $2\nu \beta \beta$ events within the fiducial volume~(FV) in this analysis. 
For $^{136}$Xe $2\nu \beta \beta$-$0_1^+$, the presence of $\gamma$s induces a double-hump structure in the spectrum. 
The left hump is primarily formed by the continuous spectrum of two electrons, with the combined energy deposition of a single $\gamma$~(760.5 keV or 818.5 keV) and the other $\gamma$ undetected or partially detected.
The right hump is produced by the spectrum of two electrons and the energy deposition of all $\gamma$s. 
In the case of $2\nu \beta \beta$, the SS ratio below 250 keV is almost 100\%, resulting in a negligible MS event contribution.

\begin{figure}[htbp]
  \centering
  \includegraphics[width=\columnwidth]{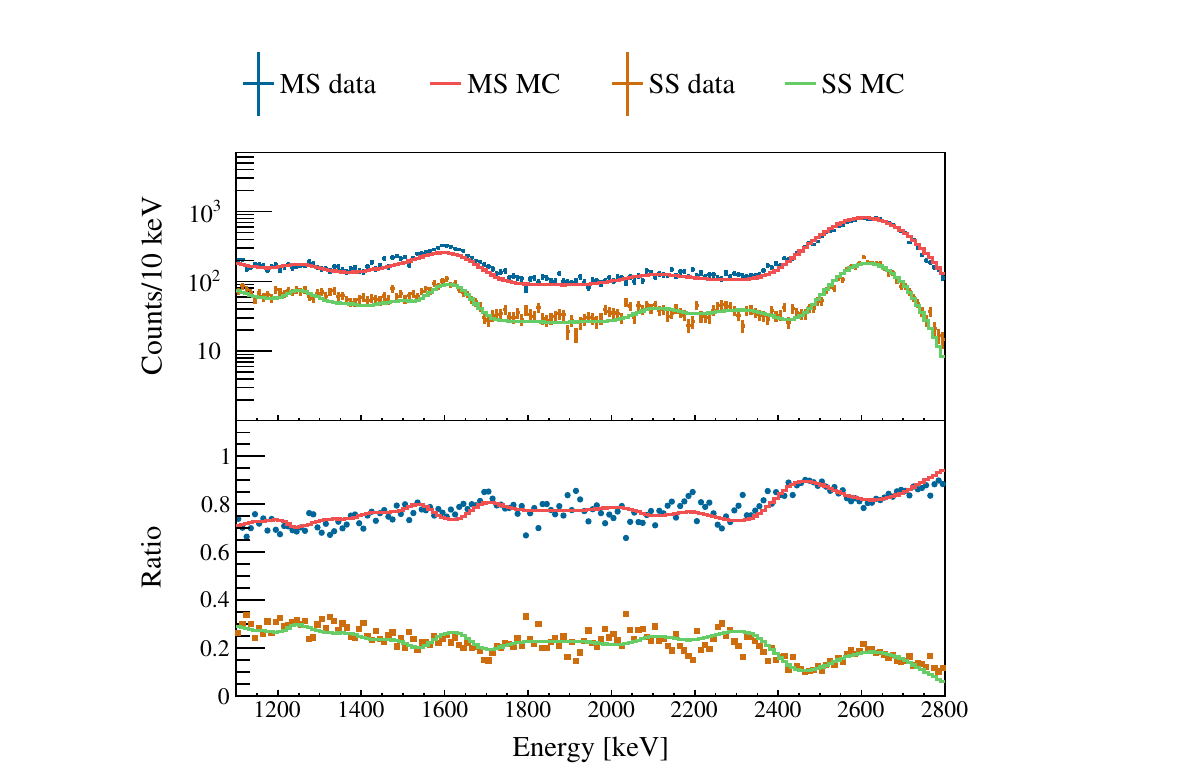}
  \caption{Energy spectra of $^{232}$Th calibration data and the Monte Carlo simulation for SS and MS events (top). The ratios of SS and MS events (bottom).}
  \label{fig:msss_agreement}
\end{figure}

The systematic uncertainty in the SS and MS ratios is evaluated using the calibration data of $^{232}$Th. 
The $^{232}$Th calibration source was placed in an external loop surrounding the detector. 
The decay chain of $^{232}$Th produces abundant $\gamma$ rays that span the entire ROI. 
As shown in Figure~\ref{fig:msss_agreement}, the relative difference in the MS ratio between data and MC, averaged over the $|$(data-MC)/MC$|$ of each bin weighted by the number of events within the ROI, is 3.4\%. 
Meanwhile, the difference in the SS ratio is 13.3\%. 
These differences are conservatively adopted as the systematic uncertainties in the simultaneous fit of the MS and SS spectra.

\subsection{Event selection}

The physical event selection criteria follow the previous analysis but are adjusted for this study to remove surface backgrounds, accidentally paired events, and events with poorly reconstructed waveforms.  
We first apply the data quality selection cuts~\cite{PandaX:2023ggs} to remove non-physical events and select electron recoil events with an efficiency of $ (99.7 \pm 0.3)\%$ for MS events and $(99.1 \pm 0.8)\%$ for SS events.
Subsequently, a vast majority of background events are rejected through ROI cut and FV cut. 
The intrinsic backgrounds in liquid xenon are mainly concentrated below 1~MeV, such as those from $^{214}$Pb. 
Therefore, the ROI is selected to range from 1100~keV to 2800~keV, preserving 99.6\% (98.6\%) of the MS (SS) 2$\nu \beta \beta$-$0^+_1$ signals.
For the FV selection, we maintain consistency with the previous SS analysis for the radial direction $\text{R}^2$, leading to a radius cut of 34.2~cm~\cite{PandaX:2022kwg}. 
The interval in the Z direction is optimized to minimize background contribution based on the MS events distribution, spanning from -901~mm to -399~mm relative to the gate grid located at the top.
This geometry corresponds to 563.40~kg of natural liquid xenon, containing 48.36~kg of $^{136}$Xe, resulting in a total exposure of 12.57~kg$\cdot$yr.

\subsection{Event reconstruction}
\label{Event recons}

The event reconstruction process includes both position and energy reconstruction. The three-dimensional position reconstruction involves determining the vertical (denoted as Z) position and the horizontal position in the X-Y plane. The Z position is determined by measuring the time difference between the S1 and S2 signals, while accounting for the drift velocity of the ionization electrons within the detector. 
The horizontal position is determined by comparing the observed S${2_\text{T}}$ charge pattern distribution in the X-Y plane with the photon acceptance function, which is derived from optical MC simulations~\cite{PANDA-X:2021jua}. 
The S${2_\text{T}}$ signal is desaturated by making waveform corrections to PMT channels with charge greater than 900 photoelectrons~\cite{Luo:2023ebw} to improve the accuracy of position reconstruction. Due to the suppression of subsequent S$2$ signals by the preceding S$2$~\cite{Luo:2023ebw}, the first peak of the S${2_\text{T}}$ signal from MS events is used for position reconstruction.

The energy reconstruction for MS and SS events is performed separately, although they share several similar procedures. The energy is calculated from S$1$ and S${2_\text{B}}$ as follows:
\begin{equation}
\begin{aligned}
  E= 13.7~\text{eV} \times  (\frac{\text{qS}1}{\text{PDE}} +\frac{\text{qS}{2_\text{B}}}{\text{EEE}\times \text{SEG}_\text{B}}).
\end{aligned}
\label{eq:E}
\end{equation}
qS$1$ and qS${2_\text{B}}$ represent the total charges of the S$1$ and S${2_\text{B}}$ signals, respectively, measured in units of photoelectrons. PDE, EEE, and SEG$_\text{B}$ denote the photon detection efficiency for S$1$, the electron extraction efficiency, and the single-electron gain for S${2_\text{B}}$, respectively. The S${2_\text{B}}$ signal is desaturated to improve the linearity of the energy reconstruction, and the uniformity correction follows previous procedures~\cite{PandaX:2022kwg,PandaX:2023ggs}.

The only difference in the reconstruction of MS events is the so-called late-pulse suppression correction, an saturation effect observed in MS event that a large S2 signal can also distort subsequent S2 pulses.
The suppression factor to the overall S${2}$ is clearly charge-dependent, which is effectively extracted based on the MS full absorption peaks of the calibration sources.

MS and SS spectra are then modeled separately using the energy resolution and residual energy scale, each characterized by a distinct set of five parameters.
The energy resolution is modeled as a Gaussian function with width $\sigma(E)$, which is expressed as $\frac{\sigma(E)}{E} = \frac{a}{\sqrt{E}} + b \cdot E + c$. 
The residual energy nonlinearity is defined as $E = d \cdot \hat{E} + e$, to account for biases between the reconstructed energy ${E}$ and the true energy $\hat{E}$.
A region of physical data, adjacent to the FV, with the same Z-range as the FV and an R range extending from 34.2 cm to 41.2 cm, is selected. 
The initial parameter values and uncertainties for both SS and MS spectra are adopted from the outer region with fitting the energy peaks of 236~keV, 1332~keV, 1461~keV and 2615~keV. Therefore these pre-fitted parameters and uncertainties are uncorrelated with later fits to the FV data.

For the MS spectra, these parameters are: 
\begin{equation}
\left\{
\begin{array}{l}
a_{0_{\text{MS}}} = 0.69\pm0.12~[\sqrt{\text{keV}}] \\
b_{0_{\text{MS}}} = (1.3\pm0.3)\times10^{-5}[\text{keV}^{-1}]\\
c_{0_{\text{MS}}} = (1.1\pm0.8)\times10^{-2}\\
d_{0_{\text{MS}}} = (1.0\pm0.0)\times10^{-3}\\
e_{0_{\text{MS}}} = -4.2\pm0.4~[\text{keV}]\\
\end{array}
\right\}.
\label{eq::MSfivepars}
\end{equation}

For SS spectra, these parameters are:
\begin{equation}
\left\{
\begin{array}{l}
a_{0_{\text{SS}}} = 0.79\pm0.22[\sqrt{\text{keV}}] \\
b_{0_{\text{SS}}} = (6.2\pm4.7)\times10^{-6}[\text{keV}^{-1}]\\
c_{0_{\text{SS}}} = (1.8\pm1.5)\times10^{-2}\\
d_{0_{\text{SS}}} = (1.0\pm0.0)\times10^{-3}\\
e_{0_{\text{SS}}} = -1.9\pm0.4~[\text{keV}]\\
\end{array}
\right\}.
\label{eq::SSfivepars}
\end{equation}

The extracted values $\mathcal{M}_0^{k}={(a_{0_{\text{k}}}, b_{0_{\text{k}}}, c_{0_{\text{k}}}, d_{0_{\text{k}}}, e_{0_{\text{k}}})^T}$ and uncertainties of the parameters (Table~\ref{tab:sys_err}) will be used as priors, together with the $5 \times 5$ covariance matrix $\Sigma_m^{k}$ in fitting the energy spectra ($k$ is the index for MS or SS). The energy resolution at 2615~keV is found to be $(3.4\pm0.3)\%$ for the MS spectrum and $(1.5\pm0.2)\%$ for the SS spectrum. 

\section{Background component}
\label{sec:bkg}

The major background contribution for $^{136}$Xe $2\nu \beta \beta$-$0_1^+$ search within ROI can be divided into three main categories: 1) $^{136}$Xe $2\nu \beta \beta$ inside the liquid xenon; 2) radioactive isotopes from the detector materials; 3) radioactives from the stainless steel platform~(SSP) surrounding the PandaX-4T cryostat.

For Category 1, the contribution from $^{136}$Xe $2\nu \beta \beta$ is calculated based on the half-life measurement in PandaX-4T~\cite{PandaX:2022kwg}. The expected number of events within the ROI is 205$\pm$10 for MS and 4839$\pm$232 for SS.
For Category 2, the activities of $^{60}$Co, $^{40}$K, $^{232}$Th, and $^{238}$U of the detector materials have been measured using high-purity germanium~(HPGe) counting stations, as reported in Ref.~\cite{PandaX-4T:2021lbm} and verified in Ref.~\cite{PandaX:2022kwg}. The background contributions from various components of the detector materials, including the flange, vessel, PMTs, and others, are weighted together based on their mass and activities.
The expected number of MS~(SS) events in the ROI is 5389 $\pm$ 3192~(663 $\pm$ 385), 6596 $\pm$ 3236~(775 $\pm$ 379),
 5382 $\pm$ 4203~(417 $\pm$ 344), and 3591 $\pm$ 2811~(336 $\pm$ 292), for $^{60}$Co, $^{40}$K, $^{232}$Th, and $^{238}$U 
respectively. 

\begin{figure}[htbp]
  \centering
  \includegraphics[scale=0.35]{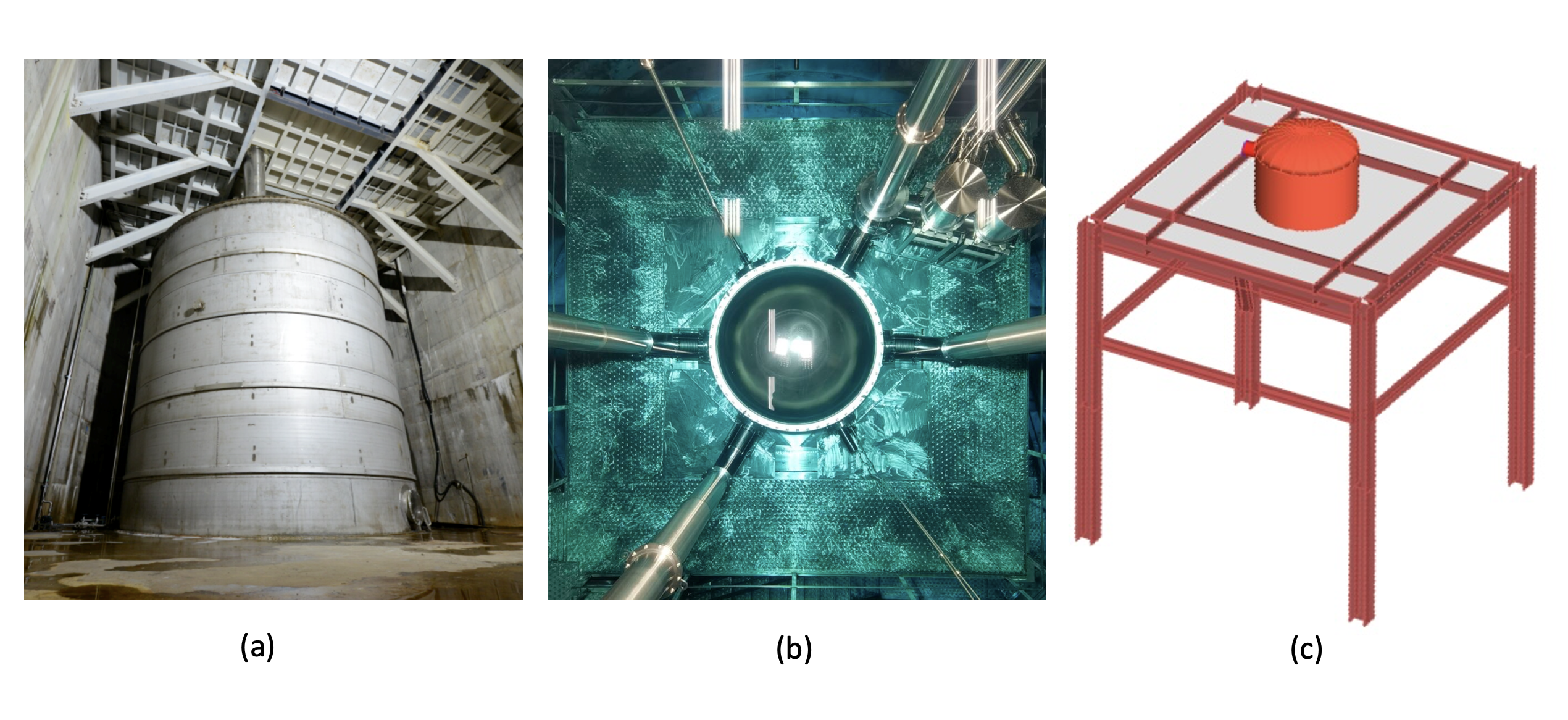}
  \caption{(a) The water tank for the PandaX-4T detector. (b) An aerial view of the detector and stainless steel platform submerged in the water. (c) The geometric model of stainless steel platform in our simulation.}
  \label{fig:steelplatform}
\end{figure}

The usage of MS events as our main signal of this analysis allows us to identify a new component of background originated from the stainless steel platform~(Category 3). The stainless steel platform, located outside the PandaX-4T cryostat and immersed inside the sheilding water tank as shown in Figure~\ref{fig:steelplatform}, is made of regular stainless steel with a total weight of more than 6.8 tonnes. Based on limited sampling, these stainless steel materials exhibit radioactivities that are 1 to 3 orders of magnitude higher than those used to make the detector vessels. Since the high-energy $\gamma$ rays from the stainless steel platform cannot be fully shielded by the water, this background component, particularly in the MeV-scale MS spectrum, becomes significant in this analysis. On the other hand, due to the shielding effects, the background spectra due to stainless steel platform exhibit a different shape in comparison to regular detector materials, which allows us to separate them \textit{in situ} with our data.

The BambooMC framework is employed to model the experimental setup, including the detector, water shielding, and stainless steel platform. The contribution of all background sources is modeled separately for SS and MS events with BambooMC and processed with the data production pipeline. As illustrated in Figure~\ref{fig:steelplatform} (c), an simplified stainless steel platform in the simulation comprises of the platform~(gray), columns, 
and beams~(red), while the central cylindrical structure represents PandaX-4T cryostat. These background spectra, in combination with the detector response models, are used to fit the data in the FV. 
The activities of  $^{60}$Co, $^{40}$K, $^{232}$Th, and $^{238}$U, are assumed to be float and uniformly distributed throughout the entire stainless steel platform. The uncertainty introduced by this assumption will be discussed in Sec.~\ref{sec:bkg fit}.

\section{Systematic uncertainties and fitting method}
\label{sec:fit method}

Before we discuss the spectral fit, let us first discuss components of systematic uncertainties, summarized in Table~\ref{tab:sys_err}, with their origins attributed to overall efficiency, detector response, signal selection, and the background model.
First, the overall efficiency encompasses factors such as the discrimination between MS and SS events and the data quality cuts, which affect the overall rate of the signal and the background.
Second, uncertainties in the detector response, as detailed in Section~\ref{sec:data}, contribute to the systematic uncertainties in the energy spectra of the signal and background.
Third, uncertainties in signal selection arise from parameters including the liquid xenon density, the $^{136}$Xe abundance, and the uniformity of the FV, which affect the rate of the signal only.
Finally, the background model introduces additional uncertainties in the rates of individual background components.

The initial values and uncertainties of these systematic components are pre-determined and incorporated into the fit to the FV data. 
The uncertainty in the MS (SS) fraction has been introduced earlier in the text.
The uncertainty in the data quality cut is 0.26\% for MS events and 0.75\% for SS events within the ROI, estimated based on variations across different detector volumes in $\text{R}^2$.
The detector response is characterized by five parameters, with their mean values $\mathcal{M}_0^k$ provided in Eq.\ref{eq::MSfivepars} and Eq.\ref{eq::SSfivepars}, and their error covariance matrix denoted as $\Sigma_m^k$.
The uncertainty in FV uniformity is estimated from the geometric distribution of \(^{83\text{m}}\)Kr, yielding an uncertainty of 1.6\%. 
We measure the abundance of $^{136}$Xe using the residual gas analyzer~\cite{RGA}, finding it to be ($8.58 \pm 0.11$)\%.
The background model uncertainties originate from the measurement of materials radioactivity ~\cite{PandaX-4T:2021lbm} and the measurement of \(^{136}\)Xe \(2\nu \beta \beta\) half-life~\cite{PandaX:2022kwg} for background Categories 1 and 2, respectively. 
As mentioned, the background levels due to Category 3 are set as free parameters instead.

\begin{table}[tbp]
  \caption{Summary of systematic uncertainties.}
  \centering
  \begin{tabular}{c c c c}
    \hline
    \multicolumn{2}{c}{\textbf{Sources}} & \multicolumn{2}{c}{\textbf{Values}} \\
    \hline
    \multirow{2}{*}{\begin{tabular}[c]{@{}c@{}}\ \textbf{Overall efficiency} \end{tabular}} 
            & MS/SS fraction & 3.4\%(MS)  &13.3\%(SS)\\
   
    & Quality cut & 0.26\%(MS) & 0.75\%(SS)\\
    \hline
    \multirow{2}{*}{\begin{tabular}[c]{@{}c@{}}\ \textbf{Detector response} \end{tabular}} & Energy scale 
    & \multirow{2}{*}{\begin{tabular}[c]{@{}c@{}} \ {$\Sigma_0^\text{{MS}}$}  
    \end{tabular}} & \multirow{2}{*}{\begin{tabular}[c]{@{}c@{}} \ {$\Sigma_0^\text{{SS}}$}\cite{PandaX:2024kjp} 
    \end{tabular}} \\
            & Energy resolution & \\
    \hline
    \multirow{3}{*}{\begin{tabular}[c]{@{}c@{}}\ \textbf{Signal selection} \end{tabular}} & LXe density & \multicolumn{2}{c}{0.13\%}\\
  
            & FV uniformity & \multicolumn{2}{c}{1.63\%}\\
   
            & $^{136}$Xe abundance & \multicolumn{2}{c}{1.33\%}  \\
    \hline 
    
    \textbf{Background model} & & \multicolumn{2}{c}{Listed in Section~\ref{sec:bkg}} \\
     \hline
  \end{tabular}
  \label{tab:sys_err}
\end{table}
A binned likelihood is constructed as shown in Eq.~\ref{eq::likelihood} following the formalism in Ref.~\cite{PandaX:2024kjp},
to simultaneously fit the energy spectra of both MS and SS events. All systematic uncertainty contributions discussed above are treated as nuisance parameters.
\begin{equation}
\begin{aligned}
    L = & \prod_{k}^{\text{SS,MS}} \prod_{i=1}^{N_{bins}} \frac{(\Bar{N}_i^k)^{N_{i}^k}}{N_{i}^k!}e^{-\Bar{N}_i^k}
    \cdot \mathcal{G}(\mathcal{M}^k; \mathcal{M}^k_0, \Sigma_m^k)
    \cdot 
    \prod_{j} G(\eta_j; 0, \sigma_j),
\end{aligned}
\label{eq::likelihood}
\end{equation}
$\Bar{N}_i^k$ and $N_i^k$ denote the expected and observed numbers of events in the $i^{\text{th}}$ energy bin, respectively, and $\Bar{N}_i^k$ is defined as follows:
\begin{equation}
\begin{aligned}
   \Bar{N}_i^k = & (1+\eta^k_{eff}) \cdot [ (1+\eta_s) \cdot n^k_{s} \cdot S^k_{i} + \sum_{b=1}^{N_{bkg}} (1+\eta^k_b) \cdot n^k_{b} \cdot B^k_{b, i}],
\end{aligned}
\label{eq::Ni}
\end{equation}
The Gaussian penalty term, $\mathcal{G}(\mathcal{M}^k; \mathcal{M}^k_0, \Sigma_m^k)$, represents the energy response and includes the five parameters $\mathcal{M}_0^k$ as listed in Section~\ref{sec:data} and the covariance matrix $\Sigma_m^k$. 
The Gaussian penalty terms $G(\eta_j; 0, \sigma_j)$ constrain the parameters $\eta^k_{eff}$, $\eta_{s}$ and $\eta^k_{b}$, which are associated with uncertainties in the overall efficiency~(Table~\ref{tab:sys_err}), signal selection~(Table~\ref{tab:sys_err}),
and background model~(Section~\ref{sec:bkg}). 
$n_s^k$ and $n_b^k$ represent the counts for the signal $s$ and the background component $b$, respectively. $S_i^k$ and $B_{b, i}^k$ denote the normalized energy spectra in the $i^{\text{th}}$ bin, after accounting for the detector response.
The MS and SS spectra share the same parameters for signal and background event rates. The fraction of MS and SS events for the signal and background components is derived from Monte Carlo simulation, with a global relative systematic uncertainty of 3.4\% and 13.3\% (conservatively assumed to be uncorrelated between MS and SS), respectively.

\section{Results}
\subsection{Background only fits}
\label{sec:bkg fit}

The MS energy spectrum provides critical insight into the background model. A background-only simultaneous fit of the MS and SS spectra is performed before the signal fit. 

\begin{figure}[htbp]
  \centering
\includegraphics[width=\columnwidth]{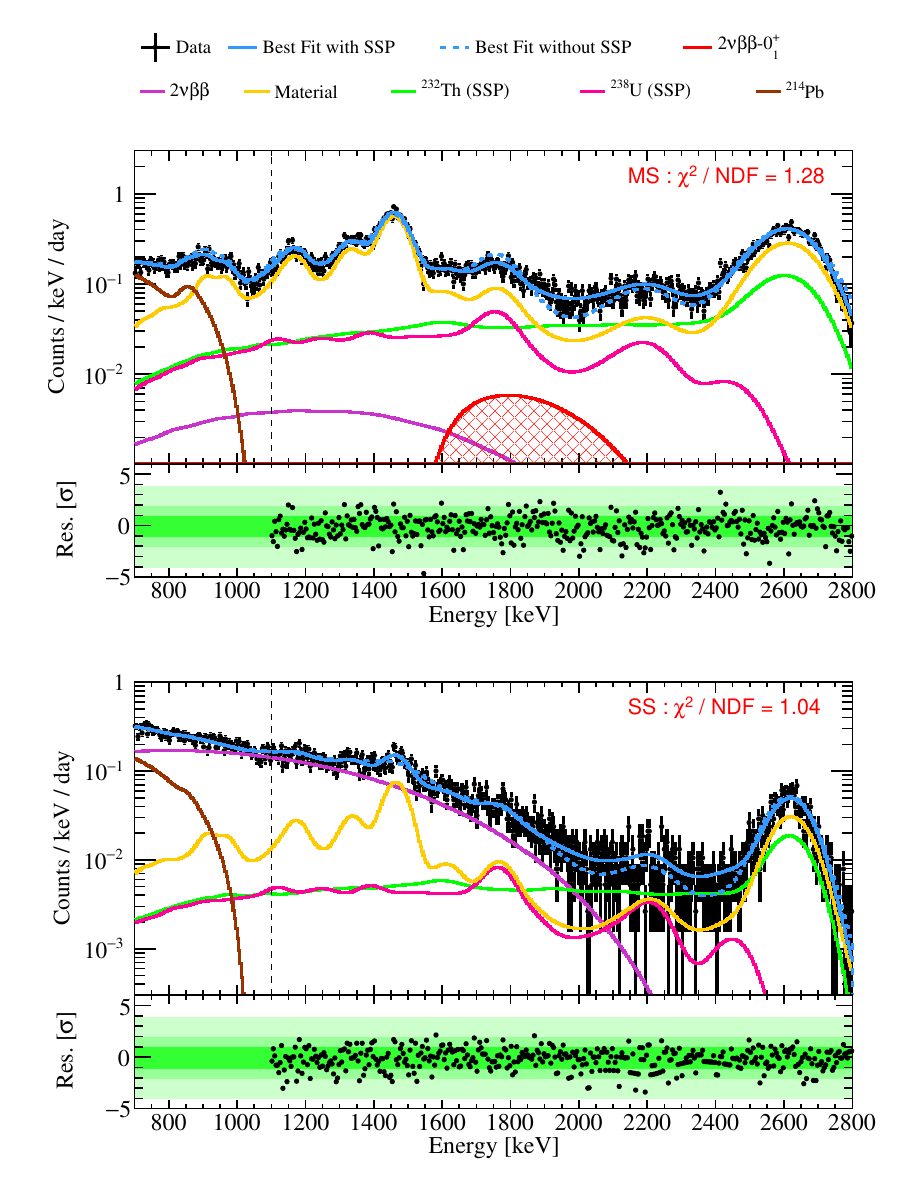}
  \caption{Background only fits to MS~(top) and SS~(bottom) data from 1100~keV to 2800~keV with bin size of 4~keV. For the sake of clarity, we have combined $^{60}$Co, $^{40}$K, $^{232}$Th, and $^{238}$U into the Material in the plot. The blue dashed line represents the best fit result excluding the stainless steel platform. 
 The energy spectrum is extended to 700 keV, as shown in the left of the black dash line. The contribution of $^{214}$Pb is from the previous analysis~\cite{PandaX:2023ggs}.
 For comparison, the final best fit $^{136}$Xe $2\nu \beta \beta$-$0_1^+$ signal is overlaid as the red hatched histogram.}
  \label{fig:signal fit}
\end{figure}
\begin{figure}[htbp]
  \centering
\includegraphics[scale=0.8]{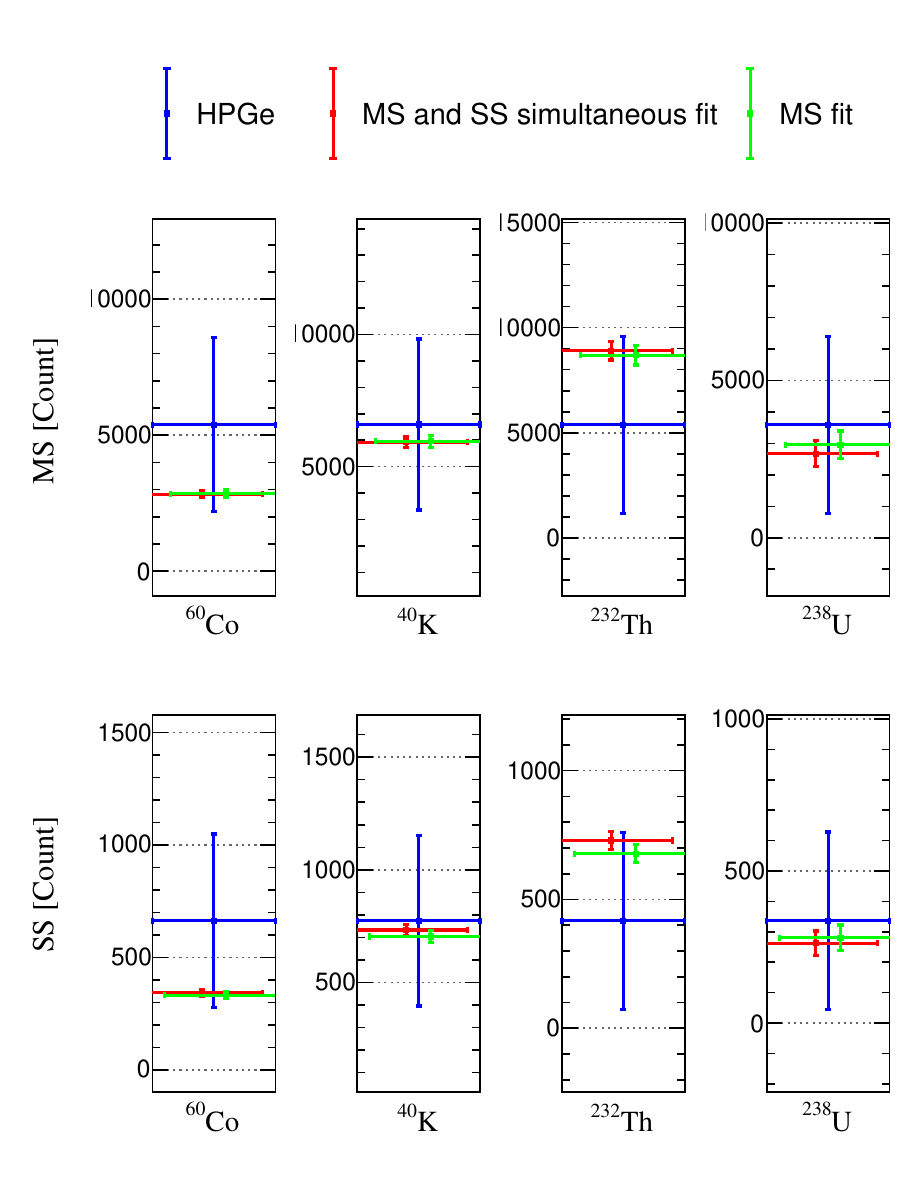}
  \caption{Background contributions of $^{60}$Co, $^{40}$K, $^{232}$Th, and $^{238}$U from detector materials.}
  \label{fig:bkg detector}
\end{figure}
\begin{figure}[htbp]
  \centering
   \includegraphics[scale=0.7]{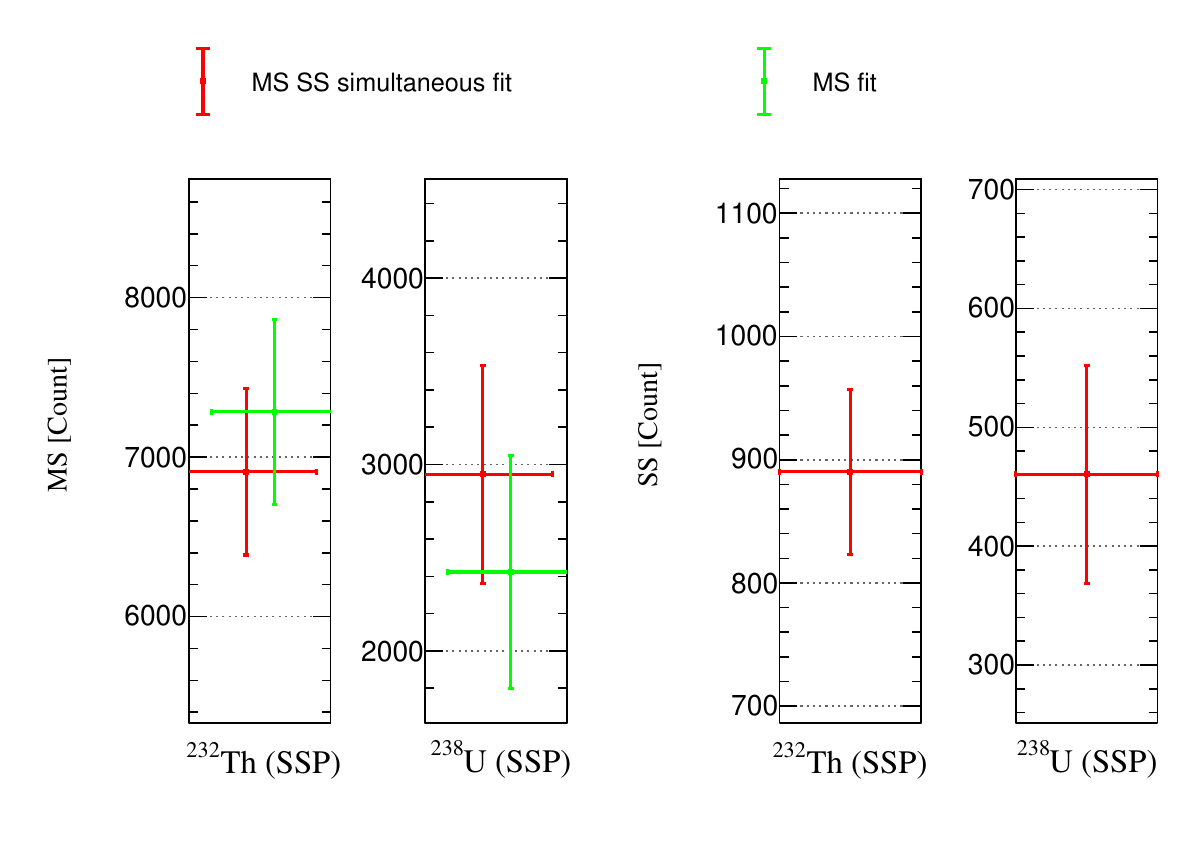}
  \caption{Background contributions of $^{232}$Th and $^{238}$U from the stainless steel platform.}
  \label{fig:bkg ssp}
\end{figure}

A fit of Category 1 and Category 2 to both MS and SS data is performed first, as shown by the blue dashed line in Figure~\ref{fig:signal fit}. The $\chi^{2}$/NDF value in the SS spectrum is 451.6/425, while in the MS spectrum, the value is 938.4/425. The 1800 keV to 2400 keV energy region shows particularly significant deviations in the MS spectrum, strongly indicating a missing background component. A fit incorporating the stainless steel platform is then performed, as shown by the blue line in Figure~\ref{fig:signal fit}. The $\chi^{2}$/NDF values for the fits decrease significantly in the MS spectrum to 543.3/423, while improving gently in the SS spectrum to 438.8/423. Clearly, the contribution of the stainless steel platform is much more pronounced in the MS spectrum than in the SS spectrum. Several conclusions can be made from the background only fit:  
\begin{itemize}
    \item The contribution of $2\nu \beta \beta$~(Category 1) from the later best background-only fit is 215 $\pm$ 8 (MS) and 5074 $\pm$ 188 (SS), which also validate that our previous results on $2\nu \beta \beta$ using only SS data is not sensitive to the inclusion of stainless steel platform.  
    \item The contributions of detector materials~(Category 2) are shown in Figure~\ref{fig:bkg detector}. The best fitted counts of MS~(SS) events for $^{60}$Co, $^{40}$K, $^{232}$Th, and $^{238}$U  is 2826 $\pm$ 128~(343 $\pm$ 156), 5919 $\pm$ 206~(733 $\pm$ 26), 8901 $\pm$ 440~(729 $\pm$ 36), and 2671 $\pm$ 410~(263 $\pm$ 40), respectively. Consistent post-fit nuissance parameters are obtained with MS-only data. The fitted contributions are consistent with the low-radioactivity material screening results reported in Ref.~\cite{PandaX-4T:2021lbm} within 1$\sigma$. This reinforces our earlier statement that MS data provides a strong constraint to the background model. 
    \item For various floating contributions from the stainless steel platform~(Category 3), for $^{60}$Co and $^{40}$K, only a narrow energy range around the energy peaks falls within the ROI. It is found that their contributions are mostly degenerate with corresponding contributions from the detector materials, therefore can be set to zero without influencing the results. The MS (SS) contributions from $^{232}$Th and $^{238}$U, on the other hand, are quite different in shape in comparison to the those from detector materials, therefore can be well-constrained by the data.
    \item The fitted MS~(SS) contributions from the stainless steel platform for $^{232}$Th and $^{238}$U are 6906 $\pm$ 520~(890 $\pm$ 67) and 2947 $\pm$ 586~(460 $\pm$ 92), respectively, as shown in Figure~\ref{fig:bkg ssp}, resulting in averaged radioactivities of \(319 \pm 24~\text{mBq/kg}\) for $^{232}$Th and \(463 \pm 92~\text{mBq/kg}\) for $^{238}$U. For comparison to lower energy data, we also draw data all the way down to 700 keV, overlaid with background model determined based on fits to data above 1100 keV. It demonstrates that below 1000 keV, the contribution of stainless steel platform is negligible particularly in the SS spectrum, and therefore has negligible impact on our previous results \cite{PandaX:2023ggs, PandaX:2024kjp} in the lower energy region.
\end{itemize}

\subsection{Signal fit}
\label{sec:results}

We have performed a simultaneous fit of the MS and SS energy spectra, incorporating the $^{136}$Xe $2\nu \beta \beta$-$0_1^+$ signals, using the same inputs and constraints as those in the background-only fit. The signal fit results are displayed in Figure~\ref{fig:signal fit} as a hatched histogram. The $\chi^{2}$/NDF is 540.7/422 and 435.2/422 for MS and SS fits. The total fitted signal number is 220 $\pm$ 124, with 217 $\pm$ 124 attributed to MS events and 3 $\pm$ 2 to SS events. The fitted nuisance parameters agree with their input values within 1$\sigma$. 
The contributions from $^{232}$Th and $^{238}$U in the stainless steel platform, as well as the $^{238}$U component from the detector materials, are reduced by 217, 232, and 53, respectively, in comparison to the background-only fit, but still well within fit uncertainties. An anti-correlation is observed between the signal and the floating stainless steel platform background. This is expected, which calls for a reduction and prior constraint of the stainless steel platform background when opportunities arise in the future.

To address the systematic uncertainty due to our assumption of uniformly distributed radioactivity on the stainless steel platform, we varied the location of the $^{232}$Th and $^{238}$U sources from 20 to 100~cm away from the cryostat vessel. Note that sources located farther away are more suppressed and therefore omitted. The maximum change in the best-fit signal is 145 counts, which is taken as an estimate of the systematic uncertainty of the stainless steel platform.

In combination with the stainless steel platform systematic uncertainty, the final best fit of $^{136}$Xe $2\nu \beta \beta$-$0_1^+$ is 220 $\pm$ 191. The $p$ values for null signal is found to be 0.13, showing that there is no statistically significant evidence for nonzero signal. The lower limit on half-life of $^{136}$Xe $2\nu \beta \beta$-$0_1^+$ is achieved as 7.5$\times$10$^{22}$ yr at the 90\% confidence leve. Our result is the first experimental limit on such a rare decay in a natural xenon-based detector, demonstrating the potential of $^{136}$Xe $2\nu \beta \beta$-$0_1^+$ search with next-generation multi-ten-tonne natural xenon detectors.

\section{Summary}
\label{sec:conclusion}

In summary, we report on the search for the double beta decay of $^{136}$Xe to the excited state $0_1^+$ of $^{136}$Ba using PandaX-4T commissioning data. 
Benefiting from the multi-site event reconstruction at the MeV scale, a more accurate background model is obtained through the data fits. The background contribution of the detector materials is determined more precisely through \textit{in situ} measurements~\cite{PandaX-4T:2021lbm, PandaX:2023ggs, PandaX:2022kwg}. More importantly, the contribution of the stainless steel platform outside the PandaX-4T cryostat is identified and evaluated for the first time, significantly contributing to the multi-site background for this particular analysis. Although the stainless steel platform background has negligible impact to our previous single-site analysis~\cite{PandaX:2023ggs, PandaX:2024kjp}, it will contribute significantly to the analysis of $^{136}$Xe neutrinoless double beta decay~\cite{PandaX:2024fed}.

A simultaneous fit of the multi-site and single-site spectra is performed, using a total exposure of $^{136}$Xe is 12.57~kg$\cdot$yr. 
No significant excess of signal over the background is observed. The lower limit on half-life for $^{136}$Xe double beta decay to the excited state $0_1^+$ of $^{136}$Ba achieved as 7.5$\times$10$^{22}$ yr at the 90\% confidence level. The limit is consistent with theoretical predictions~\cite{Jokiniemi:2022yfr} and published experimental results~\cite{EXO-200:2023pdl, KamLAND-Zen:2015tnh}. This represents the first half-life limit for such a rare decay using a natural xenon dark matter detector.
PandaX-4T has recently upgraded the ultrapure water shield into an active veto system, which will help to better constrain and suppress the stainless steel platform background. The sensitivity to such search is expected to improve in the near future. Direct assay or measures to reduce the background of stainless steel platform will be carried out when future opportunities arise.

\acknowledgments
\input{acknowledgement_20240515.tex}

\bibliographystyle{JHEP}
\bibliography{P4Xe136DBDes}

\end{document}

%% file: authorlist1002.tex
\def\shKeyLab{School of Physics and Astronomy, Shanghai Jiao Tong University, Key Laboratory for Particle Astrophysics and Cosmology (MoE), Shanghai Key Laboratory for Particle Physics and Cosmology, Shanghai 200240, China}
\def\scKeyLab{Jinping Deep Underground Frontier Science and Dark Matter Key Laboratory of Sichuan Province}
\def\BUAA{School of Physics, Beihang University, Beijing 102206, China}
\def\BUAACenter{Peng Huanwu Collaborative Center for Research and Education, Beihang University, Beijing 100191, China}
\def\BUAALab{International Research Center for Nuclei and Particles in the Cosmos \& Beijing Key Laboratory of Advanced Nuclear Materials and Physics, Beihang University, Beijing 100191, China}
\def\SCNT{Southern Center for Nuclear-Science Theory (SCNT), Institute of Modern Physics, Chinese Academy of Sciences, Huizhou 516000, China}
\def\USTClab{State Key Laboratory of Particle Detection and Electronics, University of Science and Technology of China, Hefei 230026, China}
\def\USTCdep{Department of Modern Physics, University of Science and Technology of China, Hefei 230026, China}
\def\pku{State Key Laboratory of Nuclear Physics and Technology, School of Physics,Peking University, Beijing 100871, China.}
\def\YaLongSD{Yalong River Hydropower Development Company, Ltd., 288 Shuanglin Road, Chengdu 610051, China}
\def\TDLee{New Cornerstone Science Laboratory, Tsung-Dao Lee Institute, Shanghai Jiao Tong University, Shanghai 201210, China}
\def\MESJTU{School of Mechanical Engineering, Shanghai Jiao Tong University, Shanghai 200240, China}
\def\SYU{School of Physics, Sun Yat-Sen University, Guangzhou 510275, China}
\def\SYUSFI{Sino-French Institute of Nuclear Engineering and Technology, Sun Yat-Sen University, Zhuhai, 519082, China}
\def\NKU{School of Physics, Nankai University, Tianjin 300071, China}
\def\YTU{Department of Physics, Yantai University, Yantai 264005, China}
\def\FDU{Key Laboratory of Nuclear Physics and Ion-beam Application (MOE), Institute of Modern Physics, Fudan University, Shanghai 200433, China}
\def\SPEIT{SJTU Paris Elite Institute of Technology, Shanghai Jiao Tong University, Shanghai, 200240, China}
\def\SJTUSC{Shanghai Jiao Tong University Sichuan Research Institute, Chengdu 610213, China}
\def\SYSUzhuhai{School of Physics and Astronomy, Sun Yat-Sen University, Zhuhai 519082, China}
\def\CDUT{College of Nuclear Technology and Automation Engineering, Chengdu University of Technology, Chengdu 610059, China}
\def\SDUdep{Research Center for Particle Science and Technology, Institute of Frontier and Interdisciplinary Science, Shandong University, Qingdao 266237, Shandong, China}
\def\SDUlab{Key Laboratory of Particle Physics and Particle Irradiation of Ministry of Education, Shandong University, Qingdao 266237, Shandong, China}
\def\UMD{Department of Physics, University of Maryland, College Park, Maryland 20742, USA}

\author[b]{Lingyin Luo} 
\author[c]{Zihao Bo} 
\author[c]{Wei Chen}
\author[c,a,d]{Xun Chen} 
\author[e,d]{Yunhua Chen}
\author[f]{Zhaokan Cheng}
\author[a]{Xiangyi Cui} 
\author[g]{Yingjie Fan}
\author[h]{Deqing Fang}
\author[c]{Zhixing Gao}
% ...

\affiliation[a]{\TDLee}
\affiliation[b]{\pku}
\affiliation[c]{\shKeyLab}
\affiliation[d]{\scKeyLab}
\affiliation[e]{\YaLongSD}
\affiliation[f]{\SYUSFI}
\affiliation[g]{\YTU}
\affiliation[h]{\FDU}
% ...

\emailAdd{shaobo.wang@sjtu.edu.cn}

\collaboration{PandaX Collaboration}

%% file: acknowledgement_20240515.tex


This project is supported in part by grants from National Science Foundation of China (Nos. 12090060, 12090061, 12305121, U23B2070), a grant from the Ministry of Science and Technology of China (Nos. 2023YFA1606200, 2023YFA1606201, 2023YFA1606202, 2023YFA1606204), China Postdoctoral Science Foundation (No. 2023M744093), and Office of Science and Technology, Shanghai Municipal Government (grant No. 22JC1410100, 21TQ1400218).
We thank for the support by the Fundamental Research Funds for the Central Universities. We also thank the sponsorship from the Chinese Academy of Sciences Center for Excellence in Particle Physics (CCEPP), Hongwen Foundation in Hong Kong, New Cornerstone Science Foundation, Tencent Foundation in China, and Yangyang Development Fund. Finally, we thank the CJPL administration and the Yalong River Hydropower Development Company Ltd. for indispensable logistical support and other help. 

%% file: draft_JHEP.bbl
\providecommand{\href}[2]{#2}\begingroup\raggedright\begin{thebibliography}{10}

\bibitem{Goeppert-Mayer:1935uil}
M.~Goeppert-Mayer, \emph{{Double Beta-disintegration}},
  \href{https://doi.org/10.1103/PhysRev.48.512}{\emph{Phys. Rev.} {\bfseries
  48} (1935) 512}.

\bibitem{Saakyan:2013yna}
R.~Saakyan, \emph{{Two-Neutrino Double-Beta Decay}},
  \href{https://doi.org/10.1146/annurev-nucl-102711-094904}{\emph{Ann. Rev.
  Nucl. Part. Sci.} {\bfseries 63} (2013) 503}.

\bibitem{Furry:1939qr}
W.H.~Furry, \emph{{On transition probabilities in double beta-disintegration}},
  \href{https://doi.org/10.1103/PhysRev.56.1184}{\emph{Phys. Rev.} {\bfseries
  56} (1939) 1184}.

\bibitem{Avignone:2007fu}
F.T.~Avignone, III, S.R.~Elliott and J.~Engel, \emph{{Double Beta Decay,
  Majorana Neutrinos, and Neutrino Mass}},
  \href{https://doi.org/10.1103/RevModPhys.80.481}{\emph{Rev. Mod. Phys.}
  {\bfseries 80} (2008) 481} [\href{https://arxiv.org/abs/0708.1033}{{\ttfamily
  0708.1033}}].

\bibitem{KamLAND-Zen:2019imh}
{\scshape KamLAND-Zen} collaboration, \emph{{Precision measurement of the
  $^{136}$Xe two-neutrino $\beta\beta$ spectrum in KamLAND-Zen and its impact
  on the quenching of nuclear matrix elements}},
  \href{https://doi.org/10.1103/PhysRevLett.122.192501}{\emph{Phys. Rev. Lett.}
  {\bfseries 122} (2019) 192501}
  [\href{https://arxiv.org/abs/1901.03871}{{\ttfamily 1901.03871}}].

\bibitem{NEXT:2021dqj}
{\scshape NEXT} collaboration, \emph{{Measurement of the $^{136}$Xe
  two-neutrino double-\ensuremath{\beta}-decay half-life via direct background
  subtraction in NEXT}},
  \href{https://doi.org/10.1103/PhysRevC.105.055501}{\emph{Phys. Rev. C}
  {\bfseries 105} (2022) 055501}
  [\href{https://arxiv.org/abs/2111.11091}{{\ttfamily 2111.11091}}].

\bibitem{PandaX:2022kwg}
{\scshape PandaX} collaboration, \emph{{Determination of Double Beta Decay
  Half-Life of $^{136}$Xe with the PandaX-4T Natural Xenon Detector}},
  \href{https://doi.org/10.34133/2022/9798721}{\emph{Research} {\bfseries 2022}
  (2022) 9798721} [\href{https://arxiv.org/abs/2205.12809}{{\ttfamily
  2205.12809}}].

\bibitem{GERDA:2023wbr}
{\scshape GERDA, (GERDA Collaboration)*} collaboration, \emph{{Final Results of
  GERDA on the Two-Neutrino Double-\ensuremath{\beta} Decay Half-Life of
  Ge76}}, \href{https://doi.org/10.1103/PhysRevLett.131.142501}{\emph{Phys.
  Rev. Lett.} {\bfseries 131} (2023) 142501}
  [\href{https://arxiv.org/abs/2308.09795}{{\ttfamily 2308.09795}}].

\bibitem{Majorana:2022udl}
{\scshape Majorana} collaboration, \emph{{Final Result of the Majorana
  Demonstrator\textquoteright{}s Search for Neutrinoless
  Double-\ensuremath{\beta} Decay in $^{76}$Ge}},
  \href{https://doi.org/10.1103/PhysRevLett.130.062501}{\emph{Phys. Rev. Lett.}
  {\bfseries 130} (2023) 062501}
  [\href{https://arxiv.org/abs/2207.07638}{{\ttfamily 2207.07638}}].

\bibitem{CUORE:2022jto}
{\scshape CUORE} collaboration, \emph{{Search for Majorana neutrinos exploiting
  millikelvin cryogenics with CUORE}},
  \href{https://doi.org/10.1038/s41586-022-04497-4}{\emph{Nature} {\bfseries
  604} (2022) 53}.

\bibitem{KamLAND-Zen:2022tow}
{\scshape KamLAND-Zen} collaboration, \emph{{Search for the Majorana Nature of
  Neutrinos in the Inverted Mass Ordering Region with KamLAND-Zen}},
  \href{https://doi.org/10.1103/PhysRevLett.130.051801}{\emph{Phys. Rev. Lett.}
  {\bfseries 130} (2023) 051801}
  [\href{https://arxiv.org/abs/2203.02139}{{\ttfamily 2203.02139}}].

\bibitem{Garfagnini:2024rvs}
{\scshape GERDA} collaboration, \emph{{Latest results from GERDA Phase II
  experiment on $^{76}$Ge double-beta decay and exotic decay searches}},
  \href{https://doi.org/10.22323/1.449.0158}{\emph{PoS} {\bfseries EPS-HEP2023}
  (2024) 158}.

\bibitem{Jokiniemi:2022yfr}
L.~Jokiniemi, B.~Romeo, C.~Brase, J.~Kotila, P.~Soriano, A.~Schwenk et~al.,
  \emph{{Two-neutrino \ensuremath{\beta}\ensuremath{\beta} decay of 136Xe to
  the first excited 0+ state in 136Ba}},
  \href{https://doi.org/10.1016/j.physletb.2023.137689}{\emph{Phys. Lett. B}
  {\bfseries 838} (2023) 137689}
  [\href{https://arxiv.org/abs/2211.03764}{{\ttfamily 2211.03764}}].

\bibitem{Barabash:2017bgb}
A.S.~Barabash, \emph{{Double beta decay to the excited states: Review}},
  \href{https://doi.org/10.1063/1.5007627}{\emph{AIP Conf. Proc.} {\bfseries
  1894} (2017) 020002} [\href{https://arxiv.org/abs/1709.06890}{{\ttfamily
  1709.06890}}].

\bibitem{Barea:2015kwa}
J.~Barea, J.~Kotila and F.~Iachello, \emph{{$0\nu\beta\beta$ and
  $2\nu\beta\beta$ nuclear matrix elements in the interacting boson model with
  isospin restoration}},
  \href{https://doi.org/10.1103/PhysRevC.91.034304}{\emph{Phys. Rev. C}
  {\bfseries 91} (2015) 034304}
  [\href{https://arxiv.org/abs/1506.08530}{{\ttfamily 1506.08530}}].

\bibitem{KamLAND-Zen:2015tnh}
{\scshape KamLAND-Zen} collaboration, \emph{{Search for double-beta decay of
  $^{136}$Xe to excited states of $^{136}$Ba with the KamLAND-Zen experiment}},
  \href{https://doi.org/10.1016/j.nuclphysa.2015.11.011}{\emph{Nucl. Phys. A}
  {\bfseries 946} (2016) 171}
  [\href{https://arxiv.org/abs/1509.03724}{{\ttfamily 1509.03724}}].

\bibitem{Barabash:1995fn}
A.S.~Barabash et~al., \emph{{Two neutrino double beta decay of Mo-100 to the
  first excited 0+ state in Ru-100}},
  \href{https://doi.org/10.1016/0370-2693(94)01657-X}{\emph{Phys. Lett. B}
  {\bfseries 345} (1995) 408}.

\bibitem{NEMO-3:2014pkc}
{\scshape NEMO-3} collaboration, \emph{{Investigation of double beta decay of
  $^{100}$Mo to excited states of $^{100}$Ru}},
  \href{https://doi.org/10.1016/j.nuclphysa.2014.01.008}{\emph{Nucl. Phys. A}
  {\bfseries 925} (2014) 25} [\href{https://arxiv.org/abs/1402.7196}{{\ttfamily
  1402.7196}}].

\bibitem{Kidd:2014hra}
M.F.~Kidd, J.H.~Esterline, S.W.~Finch and W.~Tornow, \emph{{Two-neutrino
  double-$\beta$ decay of $^{150}$Nd to excited final states in $^{150}$Sm}},
  \href{https://doi.org/10.1103/PhysRevC.90.055501}{\emph{Phys. Rev. C}
  {\bfseries 90} (2014) 055501}
  [\href{https://arxiv.org/abs/1411.3755}{{\ttfamily 1411.3755}}].

\bibitem{CUORE:2021xns}
{\scshape CUORE} collaboration, \emph{{Search for double-beta decay of $\mathrm
  {^{130}Te}$ to the $0^+$ states of $\mathrm {^{130}Xe}$ with CUORE}},
  \href{https://doi.org/10.1140/epjc/s10052-021-09317-z}{\emph{Eur. Phys. J. C}
  {\bfseries 81} (2021) 567}
  [\href{https://arxiv.org/abs/2101.10702}{{\ttfamily 2101.10702}}].

\bibitem{MAJORANA:2020shy}
{\scshape MAJORANA} collaboration, \emph{{Search for double-$\beta$ decay of
  $^{76}$Ge to excited states of $^{76}$Se with the MAJORANA DEMONSTRATOR}},
  \href{https://doi.org/10.1103/PhysRevC.103.015501}{\emph{Phys. Rev. C}
  {\bfseries 103} (2021) 015501}
  [\href{https://arxiv.org/abs/2008.06014}{{\ttfamily 2008.06014}}].

\bibitem{EXO-200:2023pdl}
{\scshape EXO-200} collaboration, \emph{{Search for two-neutrino double-beta
  decay of $^{136}$Xe to the excited state of $^{136}$Ba with the complete
  EXO-200 dataset*}},
  \href{https://doi.org/10.1088/1674-1137/aceee3}{\emph{Chin. Phys. C}
  {\bfseries 47} (2023) 103001}
  [\href{https://arxiv.org/abs/2303.01103}{{\ttfamily 2303.01103}}].

\bibitem{PandaX-4T:2021bab}
{\scshape PandaX-4T} collaboration, \emph{{Dark Matter Search Results from the
  PandaX-4T Commissioning Run}},
  \href{https://doi.org/10.1103/PhysRevLett.127.261802}{\emph{Phys. Rev. Lett.}
  {\bfseries 127} (2021) 261802}
  [\href{https://arxiv.org/abs/2107.13438}{{\ttfamily 2107.13438}}].

\bibitem{PandaX:2023ggs}
{\scshape PandaX} collaboration, \emph{{Searching for Two-Neutrino and
  Neutrinoless Double Beta Decay of Xe134 with the PandaX-4T Experiment}},
  \href{https://doi.org/10.1103/PhysRevLett.132.152502}{\emph{Phys. Rev. Lett.}
  {\bfseries 132} (2024) 152502}
  [\href{https://arxiv.org/abs/2312.15632}{{\ttfamily 2312.15632}}].

\bibitem{PandaX:2024pme}
{\scshape PandaX} collaboration, \emph{{Search for Cosmic-Ray Boosted Sub-MeV
  Dark-Matter\textendash{}Electron Scattering in PandaX-4T}},
  \href{https://doi.org/10.1103/PhysRevLett.133.101805}{\emph{Phys. Rev. Lett.}
  {\bfseries 133} (2024) 101805}
  [\href{https://arxiv.org/abs/2403.08361}{{\ttfamily 2403.08361}}].

\bibitem{PandaX:2024jjs}
{\scshape (PandaX),, PandaX} collaboration, \emph{{Measurement of solar pp
  neutrino flux using electron recoil data from PandaX-4T commissioning run*}},
  \href{https://doi.org/10.1088/1674-1137/ad582a}{\emph{Chin. Phys. C}
  {\bfseries 48} (2024) 091001}
  [\href{https://arxiv.org/abs/2401.07045}{{\ttfamily 2401.07045}}].

\bibitem{PandaX:2023xgl}
{\scshape PandaX} collaboration, \emph{{Search for
  Dark-Matter\textendash{}Nucleon Interactions with a Dark Mediator in
  PandaX-4T}},
  \href{https://doi.org/10.1103/PhysRevLett.131.191002}{\emph{Phys. Rev. Lett.}
  {\bfseries 131} (2023) 191002}
  [\href{https://arxiv.org/abs/2308.01540}{{\ttfamily 2308.01540}}].

\bibitem{Li:2014rca}
J.~Li, X.~Ji, W.~Haxton and J.S.Y.~Wang, \emph{{The second-phase development of
  the China JinPing underground Laboratory}},
  \href{https://doi.org/10.1016/j.phpro.2014.12.055}{\emph{Phys. Procedia}
  {\bfseries 61} (2015) 576} [\href{https://arxiv.org/abs/1404.2651}{{\ttfamily
  1404.2651}}].

\bibitem{PandaX-4T:2021lbm}
{\scshape PandaX-4T} collaboration, \emph{{Low radioactive material screening
  and background control for the PandaX-4T experiment}},
  \href{https://doi.org/10.1007/JHEP06(2022)147}{\emph{JHEP} {\bfseries 06}
  (2022) 147} [\href{https://arxiv.org/abs/2112.02892}{{\ttfamily
  2112.02892}}].

\bibitem{PandaX:2024kjp}
{\scshape PandaX} collaboration, \emph{{Searching for MeV-scale Axion-like
  Particles and Dark Photons with PandaX-4T}},
  \href{https://arxiv.org/abs/2409.00773}{{\ttfamily 2409.00773}}.

\bibitem{PandaX:2024xpq}
{\scshape PandaX} collaboration, \emph{{Measurement of two-neutrino double
  electron capture half-life of $^{124}$Xe with PandaX-4T}},
  \href{https://arxiv.org/abs/2411.14355}{{\ttfamily 2411.14355}}.

\bibitem{GEANT4:2002zbu}
{\scshape GEANT4} collaboration, \emph{{GEANT4--a simulation toolkit}},
  \href{https://doi.org/10.1016/S0168-9002(03)01368-8}{\emph{Nucl. Instrum.
  Meth. A} {\bfseries 506} (2003) 250}.

\bibitem{Chen:2021asx}
X.~Chen et~al., \emph{{BambooMC \textemdash{} A Geant4-based simulation program
  for the PandaX experiments}},
  \href{https://doi.org/10.1088/1748-0221/16/09/T09004}{\emph{JINST} {\bfseries
  16} (2021) T09004} [\href{https://arxiv.org/abs/2107.05935}{{\ttfamily
  2107.05935}}].

\bibitem{Ponkratenko:2000um}
O.A.~Ponkratenko, V.I.~Tretyak and Y.G.~Zdesenko, \emph{{The Event generator
  DECAY4 for simulation of double beta processes and decay of radioactive
  nuclei}}, \href{https://doi.org/10.1134/1.855784}{\emph{Phys. Atom. Nucl.}
  {\bfseries 63} (2000) 1282}
  [\href{https://arxiv.org/abs/nucl-ex/0104018}{{\ttfamily nucl-ex/0104018}}].

\bibitem{PANDA-X:2021jua}
{\scshape PANDA-X} collaboration, \emph{{Horizontal position reconstruction in
  PandaX-II}},
  \href{https://doi.org/10.1088/1748-0221/16/11/P11040}{\emph{JINST} {\bfseries
  16} (2021) P11040} [\href{https://arxiv.org/abs/2106.08380}{{\ttfamily
  2106.08380}}].

\bibitem{Luo:2023ebw}
L.~Luo et~al., \emph{{Improvement on the linearity response of PandaX-4T with
  new photomultiplier tube bases}},
  \href{https://doi.org/10.1088/1748-0221/19/05/P05021}{\emph{JINST} {\bfseries
  19} (2024) P05021} [\href{https://arxiv.org/abs/2401.00373}{{\ttfamily
  2401.00373}}].

\bibitem{RGA}
\url{https://www.hidenanalytical.com/products/residual-gas-analysis/}.

\bibitem{PandaX:2024fed}
{\scshape PandaX} collaboration, \emph{{Searching for Neutrinoless Double-Beta
  Decay of $^{136}$Xe with PandaX-4T}},
  \href{https://arxiv.org/abs/2412.13979}{{\ttfamily 2412.13979}}.

\end{thebibliography}\endgroup
